# Thomson microwave scattering for electron number density diagnostics of miniature plasmas at low pressure

*Xingxing Wang[1], Apoorv Ranjan[2], Animesh Sharma[3], Alexey Shashurin[4]*

*Purdue University, West Lafayette, IN, 47907*

*Mikhail N. Shneider[5]*

*Princeton University, Princeton, NJ, 08544*

This work proposes a novel method of Thomson microwave scattering for electron number density measurements of miniature plasmas at pressures <10 Torr. This method is applied to determine electron number density in a positive column of glow discharge initiated at 5 Torr in air with a plasma column diameter of 3.4 mm. The Thomson Microwave Scattering (TMS) system measured the electron number density to be $3.36 \cdot 10^{10}$ cm$^{-3}$. The result obtained using the TMS system was validated against the measurements made using the well-known technique of microwave quarter-wave hairpin resonator. Measurements with the hairpin resonator yielded an electron number density of $2.07 \cdot 10^{10}$ cm$^{-3}$ providing adequate agreement with the TMS system.

## I.     Nomenclature

| | | |
|---|---|---|
| $A$ | = | proportionality coefficient |
| $\alpha$ | = | volumetric fraction of the glass within the capacitor |
| $c$ | = | speed of light |
| $V$ | = | volume of the dielectric material |
| $E_i$ | = | Field strength of the incident wave |
| $e$ | = | electron charge |
| $\varepsilon_0$ | = | vacuum permittivity |
| $\varepsilon$ | = | dielectric constant of the medium |
| $\varepsilon_g$ | = | relative permittivity of glass |
| $\varepsilon_p$ | = | relative permittivity of plasma |
| $f$ | = | resonance frequency of the resonator |
| $L$ | = | length of the resonator |
| $m$ | = | mass of electron |
| $N_e$ | = | Total electron number |
| $n_e$ | = | electron number density |
| $P$ | = | dipole moment of the plasma channel |
| $s$ | = | displacement of electron |
| $U_I$ | = | In-phase output voltage from I/Q mixer |
| $U_Q$ | = | quadrature output voltage from I/Q mixer |

[1] PhD Student, School of Aeronautics and Astronautics
[2] Graduate student, School of Aeronautics and Astronautics
[3] PhD student, School of Aeronautics and Astronautics
[4] Assistant professor, School of Aeronautics and Astronautics
[5] Senior research scholar, Mechanical & Aerospace Engineering

| | | |
|---|---|---|
| $U_{out}$ | = | Total output signal voltage from I/Q mixer |
| $v_g$ | = | collisional frequency |
| $\omega$ | = | frequency of the incident wave |
| $\omega_p$ | = | plasma frequency |

## II. Introduction

The microwave scattering technique used for diagnostics of microplasmas was first proposed theoretically by Shneider and was then successfully implemented in a number of relative measurements of laser-induced avalanche ionization in air, resonance-enhanced multiphoton ionization in argon, and atmospheric pressure plasma jets [1] [2] [3] [4] [5].

Absolute measurements of the electron plasma density by means of the microwave scattering technique were proposed by Shashurin and demonstrated with several plasma objects including non-equilibrium atmospheric pressure plasma jets (APPJ), microdischarges used for electrosurgery, nanosecond repetitive pulsed discharges in air, and laser-induced plasmas [6] [7] [8] [9] [10] [11] [12] [13].

The rationale of the approach is to ensure scattering the microwave radiation off the plasma volume in the quasi-Rayleigh regime (denoted as RMS in the following description) such that the incident microwave electric field is distributed uniformly across the entire plasma volume. This is achieved when the prolonged plasma volume is oriented along the linearly polarized microwaves and the plasma diameter is small compared to the spatial scale of the microwave field ("Quasi" denotes that scattering is equivalent to classical Rayleigh scattering if incident radiation is linearly polarized along the direction of the plasma object.) In this case, the plasma electrons experience coherent oscillations in the incident microwave field and radiate a Hertzian dipole radiation pattern in a far field. Measurement of the scattered signal amplitude allows for the determination of the total number of electrons in the plasma volume and average plasma density after the appropriate calibration of the system with dielectric scatterers.

The microwave scattering technique was recently proposed for temporal measurement of electron number and electron number density. The technique is used to conduct non-intrusive direct measurements and has been implemented in a number of relative measurements of atmospheric-pressure micro-plasma samples, including atmospheric pressure plasma jet, nanosecond repetitive-pulsed discharges, micro-discharges used for electrosurgery, and laser-induced plasmas [6] [12] [13] [14]. However, microwave scattering has yet to be applied to miniature discharges at low gas pressure. In this work, measurements of electron number and electron number density in a miniature discharge at low pressure (~5 Torr) are presented. This work aims to validate electron number density for a miniature plasma at low pressures using the TMS method. To achieve this, a custom-made discharge tube at 5 Torr filled with air will be ignited by high voltage DC voltage to generate normal glow discharge. The electron number density of the positive column of the discharge will first be measured using the TMS system, which will then be compared with the density measurement obtained using a hairpin resonator inside of the plasma body.

### III. Methodology of elastic microwave scattering

The particular regime of microwave scattering is directly coupled to the type of oscillatory motion an electron experiences in the incident microwave field. Generally, the equation of motion can be written as $\ddot{s} + v_g \dot{s} + \omega_p^2 s = -\frac{e}{m} E_i e^{i\omega t}$ [6] where $s$ is electron displacement, $\omega_p$ is the plasma frequency, $E_i$ and $\omega$ are the amplitude and frequency of the incident microwave field. Assuming the case of a thin, elongated geometry of the plasma object where restoring forces can be neglected and electric field inside the plasma volume is equal to that in the incident wave, one can solve for the amplitude of electron oscillations as: $s_0 = \frac{eE_i}{m} \cdot \frac{1}{\omega\sqrt{\omega^2 + v_g^2}}$.

The Rayleigh Microwave Scattering (RMS) regime refers to the case when $v_g \gg \omega$. In this case, electrons in the plasma volume experience oscillations with an amplitude of $s_0 = \frac{eE_i}{m\omega v_g}$ due to the incident microwave field. Plasma electrons are not in free motion since they experience multiple collisions over the period of the incident microwave field oscillation. The electron collision frequency in the denominator is governed by the collisions with gas particles ($v_g$) for electron densities $<10^{17}$ cm$^{-3}$ in case of atmospheric pressure conditions. Total dipole moment of the plasma channel ($p$) can be calculated as: $p = es_0 \int n(r,z) 2\pi r dr dz = esN_e = \frac{e^2}{mv_g} \frac{E_i}{\omega} N_e$, and the amplitude of the electric field at the location of the detecting horn is correspondingly: $E_s = \frac{k^2 p}{r} = \frac{e^2}{mc^2 v_g} \frac{\omega E_i}{r} N_e$. Finally, one particularly convenient form of the expression for the output signal measured by the RMS system is $U_{out} \propto E_s \propto \frac{e^2}{mv_g} N_e$.

The Thomson Microwave Scattering (TMS) regime corresponds to the opposite case when $\omega \gg v_g$. Amplitude of electron oscillations reduces to $s_0 = \frac{eE_i}{m\omega^2}$. Plasma electrons are in free motion since no collisions occur on the period of the incident microwave field oscillation. Finally, output signal measured by the homodyne detection system can be written as $U_{out} = A \cdot \frac{e^2}{m\omega} N_e$, where $A$ is the factor of proportionality. If microwave radiation around 10 GHz is used ($\omega \sim 6 \cdot 10^{10}$ sec$^{-1}$), the TMS regime corresponds roughly to the gas pressure $P<10$ Torr since $v_g$ can be approximated as $v_g [s^{-1}] \approx 10^9 p[\text{Torr}] < 10^{10} s^{-1}$ [15].

Summarizing the above, the output signal of the microwave scattering system for the different scattering regimes can be written as follows:

$$U_{out} = \begin{cases} A \cdot \frac{e^2}{mv_g} N_e & - \text{ plasma at } p > 10 \text{ Torr (quasi} - \text{Rayleigh scattering regime)} \\ A \cdot \frac{e^2}{m\omega} N_e & - \text{ plasma at } p < 10 \text{ Torr (Thomson scattering regime)} \\ A \cdot V \cdot \varepsilon_0 (\varepsilon - 1)\omega & - \text{ dielectric scatterer} \end{cases} \quad \text{Eq. (1)}$$

## IV. Methods and Equipment

a. Custom-made discharge tube

A custom-made Pyrex glass discharge tube (dielectric constant $\varepsilon_g=4$) with an inner diameter of 3.4 mm, outer diameter of 5 mm, and an overall length of approximately 10 cm used in this experiment is shown below in Figure 1. A tungsten lead was attached at each end of the tube as the discharge electrode. A tungsten hair pin resonator with a total length of approximately 1 cm was placed in the middle of the discharge tube. A third port perpendicular to the main discharge tube was connected to a small vacuum pump. A needle valve was connected in line for minor pressure adjustments. The pressure inside of the tube was monitored by a Baratron pressure gauge.

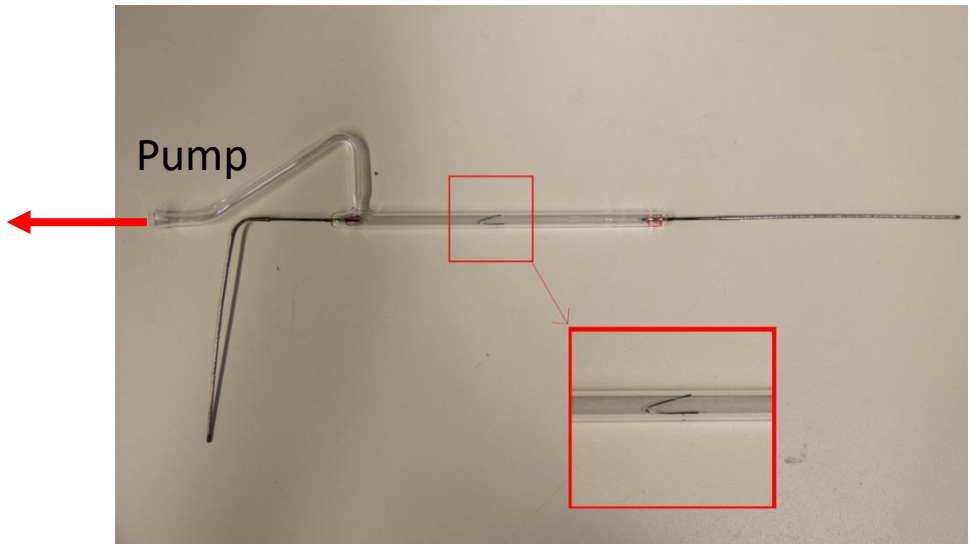

**Figure 1. Custom-made discharge tube with hairpin resonator inside**

The schematic of the electrical circuitry for the discharge tube is shown below in Figure 2. High voltage DC was supplied by a Spellman SL2000 up to 8 kV and 2 kW. A 10 kΩ resistor was connected in series on the HV side to limit the discharge current and prevent sparking.

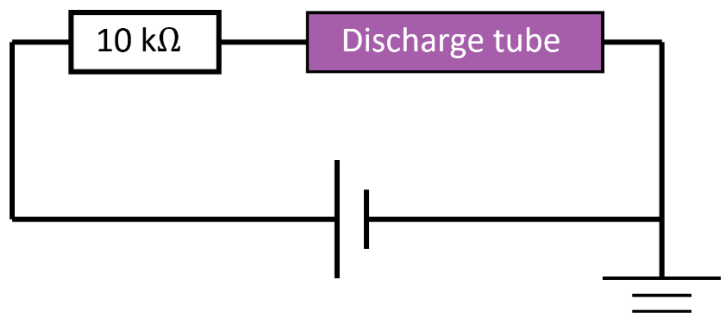

**Figure 2. Schematics of the electrical circuitry for sustaining the glow discharge**

Images taken by an ICCD camera (Princeton Instrument PI-MAX4) of the discharge tube while in operation are shown in Figure 3. With the gate width of the ICCD camera set to 100 ns, one can see from Figure 3(a) that, when the discharge tube was operating at 2 Torr and 10 mA, striations were present and the spacing between each striation was comparable to the size of the hairpin resonator [15]. By optimizing the tube pressure and discharge current, a uniformly distributed positive column was achieved at 5 Torr and 10 mA, as shown in Figure 3(b). The experiments below were conducted at these conditions to ensure uniform discharge column and absence of striations.

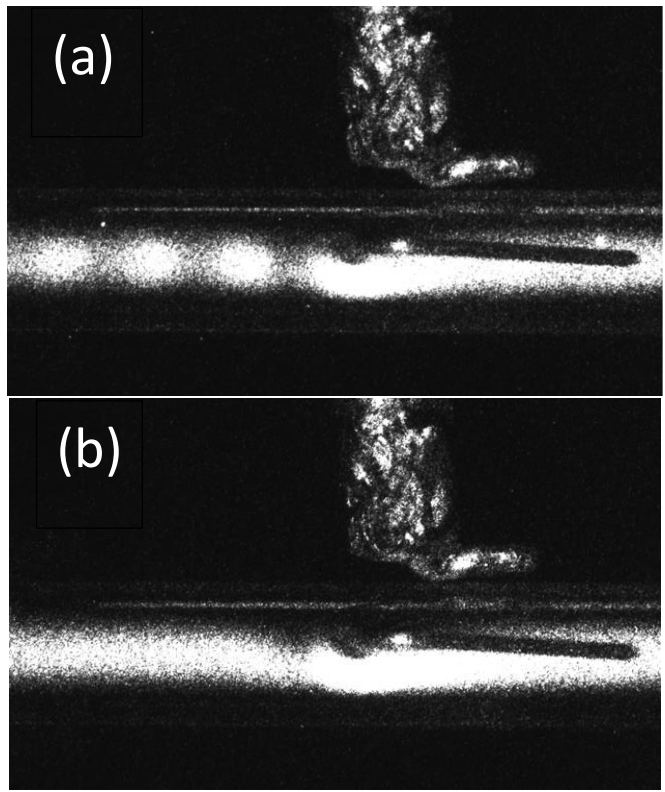

**Figure 3. ICCD images of discharge tube in operation taken at 100 ns exposure time. (a)with striations at 2 Torr. (b)without striations at 5 Torr**

b. Thomson Microwave Scattering system

A homodyne system at 10.5 GHz utilizing I/Q mixer was used for detection of the scattered signal. The schematics of the microwave circuitry were the same as those utilized in our previous studies with Rayleigh microwave scattering technique [6] [12][13]. The frequency was chosen to be 10.5 GHz. Testing objects were in the microwave field as shown schematically in Figure 4 and distance of both radiating and receiving horn antennas to the testing objects was kept at 10 cm.

Two output signals were extracted from the TMS system from the I/Q mixer: $U_I$ and $U_Q$. The final scattered signal by the object can be calculated as: $U_{out} = \sqrt{U_I^2 + U_Q^2}$.

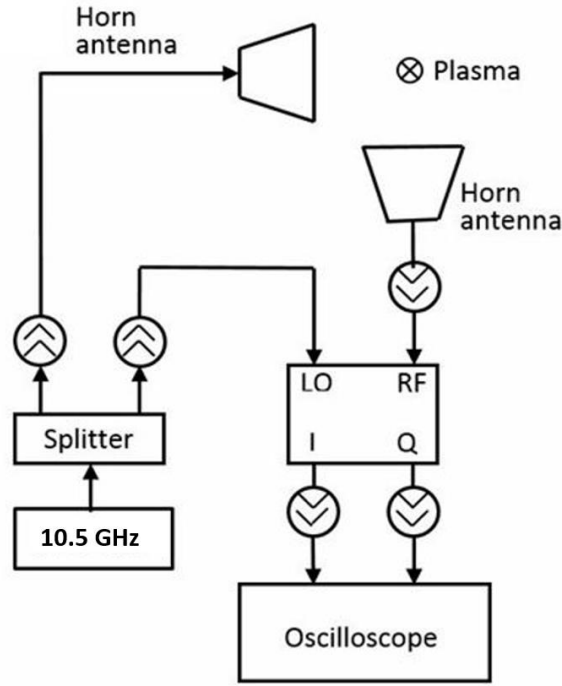

**Figure 4: Schematic of TMS system.**

Calibration of the TMS system was done prior to the measurements of scattering from the plasma. A Pyrex glass tube of same geometry and material as one used for the discharge tube, but open on both ends, was placed in the microwave field as shown schematically in Figure 4. It was then covered with microwave absorber with an opening of 2 cm at the center of the TMS detection region. A Teflon bullet with length of 2 cm and diameter of 3.2 mm (1/8 in) was driven by compressed air flying across the tube. The resultant scattered signal is shown below in Figure 5. One can see that the amplitude of the scattered signal produced by the Teflon bullet was 8.5 mV. It is worth noting that the scattered pulsed signal produced by the Teflon bullet was sitting on top of slower oscillations which are caused by the vibrations of the glass tube triggered by the application of compressed air pulse. The actual value of $U_{out}$ =8.5 mV was used to determine the calibration factor *A*. By applying the bottom relation from Eq. (1), the calibration factor *A* can be calculated as $83{,}582\,\frac{V\Omega}{m^2}$. Finally, the relation for Thomson scattering regime can be re-written as: $N_e = 2.81 \cdot 10^{13} \cdot U_{out}[V]$.

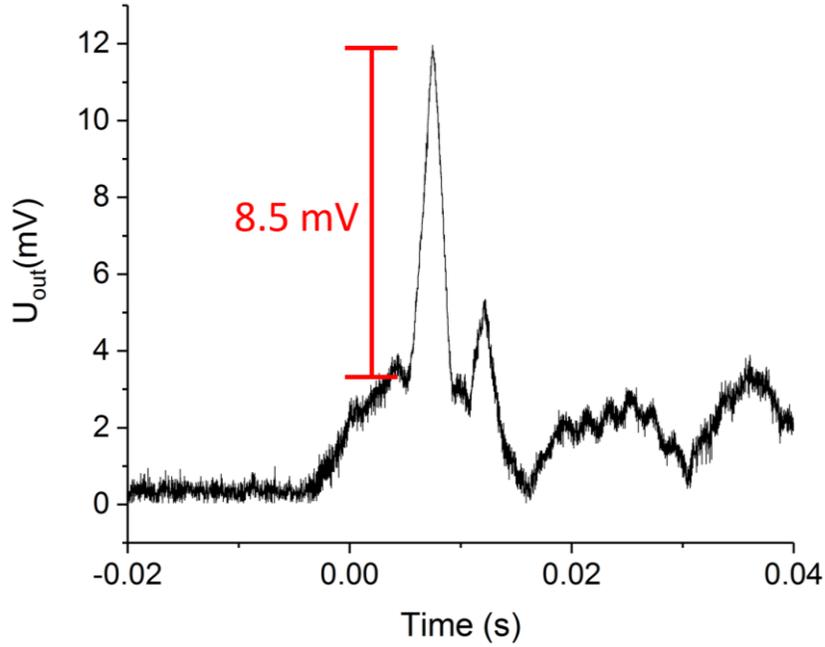

**Figure 5. Uout vs time for Teflon bullet for microwave system calibration**

When testing plasma, the discharge tube was placed at the same location as the glass tube used for calibration. A 2-cm opening of the detection region of the TMS system was created by covering the rest of the tube with microwave absorbers as shown in Figure 6. Since plasma had uniformly filled the discharge tube, the plasma channel volume was determined from the inner diameter of the glass tube (3.4 mm) and length of the exposed segment (2 cm), which yields the volume of the plasma column 0.18 cm$^{-3}$. Finally, average electron number density was determined from the scattered signal $U_{out}$ as: $n_e [cm^{-3}] = 1.56 \cdot 10^{14} \cdot U_{out}[V]$.

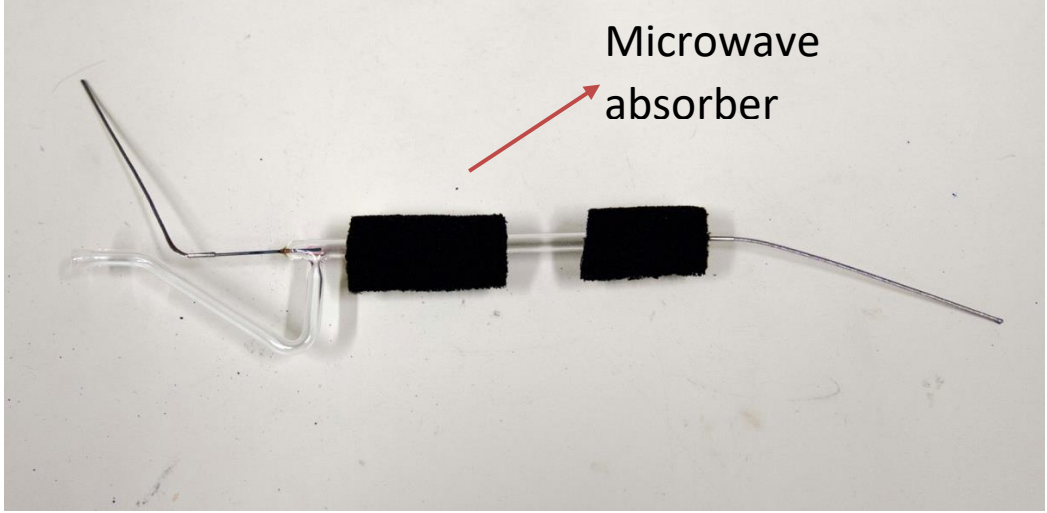

**Figure 6. Microwave absorber used to limit the length of the exposed plasma channel to 2 cm.**

c. Hairpin resonator measurement

A quarter-wave hairpin resonator was mounted inside of the discharge tube. A loop antenna placed in close proximity of the discharge tube was used to excite the resonator as shown in 6. The loop antenna was connected to a vector network analyzer (VNA) (Agilent 8722ES) and $S_{11}$ parameter was measured in a frequency range from 5 GHz to 10 GHz.

Resonance frequency of the quarter-wave hairpin resonator can be written as $f = \frac{c}{4L\sqrt{\varepsilon}}$ where $L$ – length of the resonator, $c$- speed of light, $\varepsilon$- dielectric constant of the media surrounding the resonator. Resonance frequency of the hairpin resonator was measured for three cases: (a) hairpin resonator in a free-space ($f_0$), (b) hairpin resonator installed in the discharge tube in absence of plasma ($f_g$), and (c) hairpin resonator installed in the discharge tube in presence of plasma ($f_{g+p}$). The dielectric permittivity of plasmas was determined from the resonant frequency shift due to plasma, and then plasma frequency and number density were determined using $\varepsilon_p = 1 - \frac{\omega_p^2}{\omega^2}$ and $\omega_p = 5.65 \cdot 10^4 \sqrt{n_e[\text{cm}^{-3}]}$ respectively [16]. More details are given below in Results and Discussion.

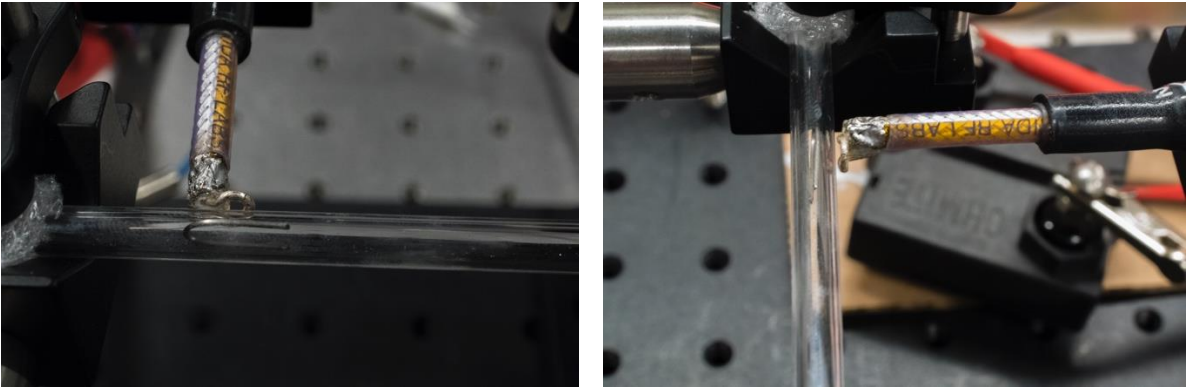

**Figure 7.** Loop antenna with hairpin resonator inside of discharge tube

## V. Results and Discussion

First, plasma density in the discharge tube was measured by TMS approach. Specifically, TMS measurements were analyzed in the vicinity of the region when the plasma was turned off in the discharge tube. This instance was associated with the step-wise change of the *I* and *Q* outputs of the mixer as shown below in Figure 8. The resultant $U_{out}$ for plasma was calculated to be 0.2154 mV. Thus, the electron number density $n_e$ measured using the TMS approach was $3.36 \cdot 10^{10}$ cm$^{-3}$.

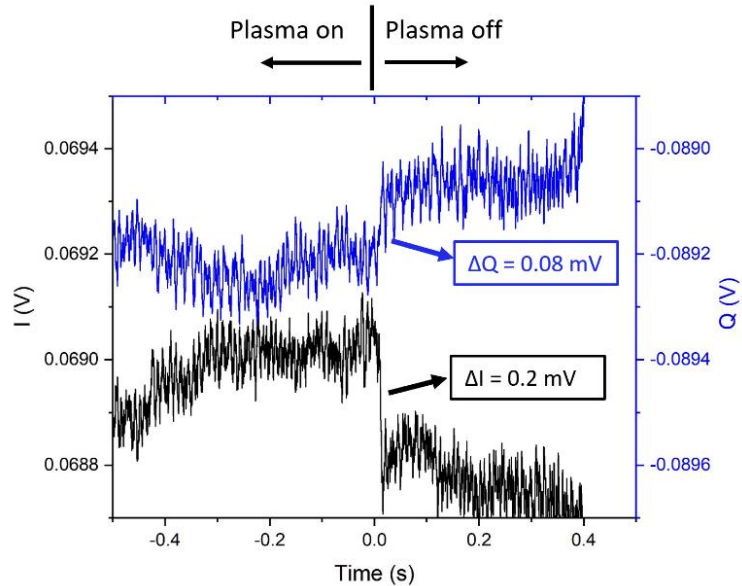

**Figure 8.** *I* and *Q* waveform at the instance when plasma was turned off

Following this, plasma density in the discharge tube was measured using the hairpin approach as follows. The resonance line of the hairpin resonator was first measured in a free-space as shown in Figure 9 (a). One can see that the resonance frequency in a free-space $f_0 = \frac{c}{4L}$ was approximately 8 GHz, which is consistent with the resonator's length $L \approx$9-10 mm used. Then, the resonance line of the hairpin resonator installed in the discharge tube (see Figure 7) was measured as shown in Figure 9 (b). One can see that the resonance frequency in presence of the tube shifted to around $f_g$ =5.64 GHz due to contribution of glass into the dielectric constant. We have approximated average dielectric constant as $\varepsilon = \alpha \cdot \varepsilon_g + (1-\alpha) \cdot 1$, where $\varepsilon_g$ - dielectric perminttivity of the glass ($\varepsilon_g$=4) and $\alpha$ can be interpreted as volumetric fraction of the space taken by glass. Based on the measured values $f_0$ and $f_g$, we have determined $\alpha$=0.33.

Finally, the resonance line of the hairpin resonator in presence of plasma in the discharge tube was measured as shown in Figure 9 (c). One can see that the resonance frequency shifted to $f_{g+p}$ =5.69 GHz due to presence of plasmas in the discharge tube. The dielectric constant of plasma ($\varepsilon_p$) was adjusted in the expression $\varepsilon = \alpha \cdot \varepsilon_g + (1-\alpha) \cdot \varepsilon_p$ in order to achieve the best fit of the resonance frequency $\frac{f_0}{\sqrt{\varepsilon}}$ to the measured value of $f_{g+p}$ =5.69 GHz. It was determined that $\varepsilon_p$= 0.9482, and thus $f_p$=1.30 GHz. Finally, average plasma density in the discharge tube was determined to be $n_e$=2.07·10$^{10}$ cm$^{-3}$.

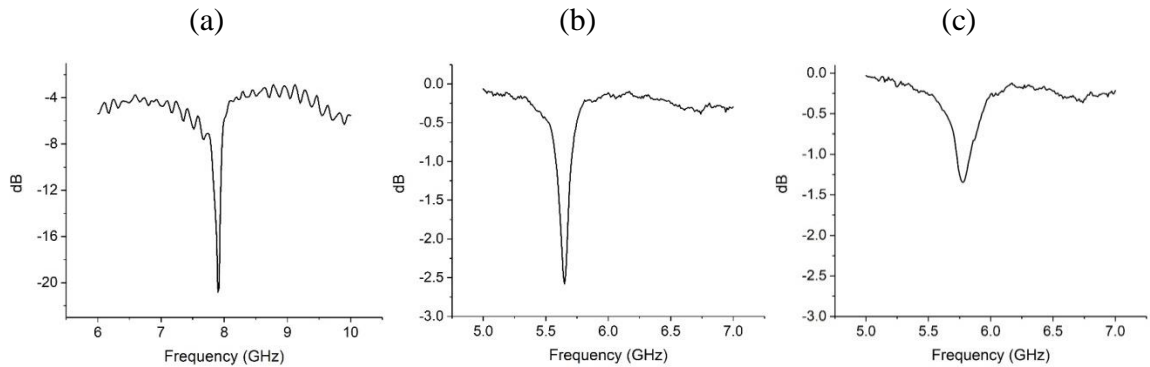

**Figure 9. Resonance line of the hairpin resonator. (a) hairpin resonator in a free-space, (b) hairpin resonator installed in the discharge tube in absence of plasma, and (c) hairpin resonator installed in the discharge tube in presence of plasma.**

Therefore, plasma density measurements conducted by the Thomson microwave scattering approach and by the hairpin technique were in fair agreement. Discrepancy of the measurements can be attributed to the perturbations inserted to the electron number density due to the presence of the hairpin resonator wire inside the discharge tube. Also, accounting for the glass discharge tube contribution to the hairpin's resonance frequency creates another possible reason for the minor difference in the results.

## VI. Conclusion

In this work, a novel approach of Thomson Microwave Scattering (TMS) to measure the electron number density for the miniature plasmas at low pressures ($p<10$ Torr) was demonstrated. The electron number density was measured by the TMS system and validated using the hairpin resonator technique. Both measurement approaches were in fair agreement; namely, TMS measurements yielded $n_e=3.36 \cdot 10^{10}$ cm$^{-3}$, while hairpin technique yielded $n_e=2.07 \cdot 10^{10}$ cm$^{-3}$.

## VII. Acknowledgements

We would like to thank Dr. S. Macheret for useful discussions, and Dr. A. Semnani for providing assistance with the hairpin measurements using the VNA.

## VIII. References

[1] M. N. Shneider and R. B. Miles, "Microwave disgnostics of small plasma objects," *J. Appl. Phys. ,* vol. 98, no. 033301, 2005.

[2] J. Sawyer , Z. Zhang and M. N. Shneider, "Microwave scattering from laser spark in air," *J. Appl. Phys,* vol. 063101, no. [15]112, 2012.

[3] A. Shashurin, M. N. Shneider, A. Dogariu, R. B. Miles and M. Keidar, "Temporal behavior of cold atmospheric plasma jet,," *Appl. Phys. Lett,* vol. 94, no. 31504, 2009.

[4] A. Dogariu, M. N. Shneider and R. B. Miles, "Versatile radar measurement of the electron loss rate in air," *Appl. Phys. Lett.,* vol. 103, no. 224102, 2013.

[5] A. Zhang, M. N. Shneider and R. B. Miles , "Microwave diagnostics of laser-induced avalanche ionization in air," *J. Appl. Phys.,* vol. 074912, no. 100, 2006.

[6] A. Sharma, M. N. Slipchenko, M. N. Shneider, X. Wang, K. A. Rahman and A. Shashurin, "Counting the electrons in a multiphoton ionization by elastic scattering of microwaves," *Scientific Reports,* vol. 8, no. 2874, 2018.

[7] A. Shashurin, "Cold Atmospheric Pressure Plasma: Technology," *Encyclopedia of Plasma Technology,* vol. 1, no. 284, 2016.

[8] A. Shahsurin and M. Keidar, "Experimental approaches for studying non-equilibrium atmospheric plasma jets," *Phys. Plasmas,* vol. 22, no. 034006, 2015.

[9] A. Shashurin, M. N. Shneider and M. Keidar, "Measurements of streamer head potential and conductivity of streamer column in cold nonequilibrium atmospheric plasmas," *Plasma Sources Sci. Technol.,* vol. 21, no. 034006, 2012.

[10] A. Shashurin, M. N. Shneider, A. Dogariu, R. B. Miles and M. Keidar, "Temporary-resolved measurement of electron density in small atmospheric plasmas," *Appl. Phys. Lett.,* vol. 96, no. 171502, 2010.

[11] A. Shashurin, D. Scott, T. Zhuang, J. Canady, I. I. Beilis and M. Keidar, "Electric discharge during electrosurgery," *Sci. Reports,* vol. 4, no. 9946, 2015.

[12] X. Wang and A. Shashurin, "Study of atmospheric pressure plasma jet parameters generated by DC voltage driven cold plasma source," *J. Appl. Phys.,* vol. 122, no. 063301, 2017.

[13] X. Wang, P. Stockett, R. Jagannath, S. Bane and A. Shashurin, "Time-Resolved Measurements of Electron Density in Nanosecond Pulsed Plasmas Using Microwave Scattering," *Plasma Source Sci. Technol.,* vol. 27, no. 07LT02, 2018.

[14] A. Shashurin, D. Scott, T. Zhuang, J. Canady, I. I. Beilis and M. Keidar, "Electric discharge during electrosurgery," *Sci. Reports,* vol. 4, no. 9946, 2015.

[15] Y. P. Raizer, Gas Discharge Physics, Berlin: Springer-Verlag, 1991, pp. 350 - 352.

[16] Y. P. Raizer, Gas Discharge Physics, Berlin: Springer-Verlag, 1991.

[17] N. L. Aleksandrove, S. B. Bodrov, M. V. Tsarev, A. A. Murzanev, Y. A. Sergeev, Y. A. Malkov and A. N. Stepanov, "Decay of femtosecond laser-induced plasma filaments in air, nitrogen, and argon for atmospheric and subatmospheric pressures," *Phys. Rev. E,* vol. 94, no. 013204, 2016.

[18] S. Bodrov, V. Bukin, M. Tsarev, A. Murzanev, S. Garnov, N. Alexksandrov and A. Stepanov, "Plasma filament investigation by transverse optical interferometry and terahertz scattering," *Optics Express,* vol. 19, no. 6829, 2011.

[19] A. A. Ovsyannikov and M. F. Zhukov, Plasma Diagnostics, Cambridge: Canbridge International Science Publishing, 2000.

[20] R. M. v. d. Horst, T. Verreycken, E. M. v. Veldhuizen and P. J. Bruggeman, "Time-resolved optical emission spectroscopy of nanosecond pulsed discharges in atmospheric-pressure N2 and N2/H2O mixtures," *J. Phys. D: Appl. Phys.,* vol. 45, no. 345201, 2012.

[21] D. L. Rusterholtz, D. A. Lacoste, G. D. Stancu, D. Z. Pai and C. O. Laux, "Ultrafast heating and oxygen disssociation in atmospheric pressure air by nanosecond repetitively pulsed discharges," *J. Appl. D: Appl. Phys,* vol. 46, 2013.

[22] F. Sainct, D. Lacoste and C. Laux, "Investigation of water dissociation by Nanosecond Repetitive Pulsed Discharge in superheated steam at atmospheric pressure," in *51th AIAA Aerospace Sciences Meeting including the New Horizons Forum and Aerospace Exposition*, Grapevine, Texas, 2013.

[23] X.-M. Zhu, J. L. Wlash, W.-C. Chen and Y.-K. Pu, "Measurement of the temperal evolution of electron density in a nanosecond pulsed argon microplasma: using both Stark broadening and an OES line-ratio method," *J. Phys. D: Appl. Phys.,* vol. 45, no. 295201, 2012.

[24] D. Pai, D. Lacoste and C. Laux, "Nanosecond reprtitively pulsed discharge in air at atmospheric pressure - spark regime," *Plasma Sources Sci. Technol.,* vol. 19, no. 065015, 2010.

[25] R. L. Stenzel, "Microwave resonator probe for localized density measrement in weakly magnetized plasmas," *Review of Scientific Intruments,* vol. 47, pp. 603-607, 1976.